\documentclass{aa} 
\usepackage{graphicx}
\usepackage{color}
\DeclareSymbolFont{cmletters}     {OML}{cmm}{m}{it}
\DeclareMathSymbol{\varB}       {\mathord}{cmletters}{`B}
\DeclareMathSymbol{\varV}       {\mathord}{cmletters}{`V}
\DeclareMathSymbol{\varR}       {\mathord}{cmletters}{`R}
\usepackage{txfonts}
\begin{document}
\title{Wide-field kinematics of globular clusters 
in the Leo I group\thanks{Based on observations collected at the very large
telescope~(VLT), European Southern Observatory, Chile, under programme 
71.B--0537.}\fnmsep\thanks{Tables 
\ref{globulars}, \ref{stars} and \ref{galaxies} are 
available in electronic~form~at CDS ({\tt ftp cdsarc.u-strasbg.fr}) 
or {\tt http://www.edpsciences.org}}} 
\author{G. Bergond\inst{1,2,3}
  \and S. E. Zepf\inst{1}
  \and A. J. Romanowsky\inst{4,5}
  \and R. M. Sharples\inst{6}
  \and K. L. Rhode\inst{7,8}}
\offprints{S. Zepf, \email{zepf@pa.msu.edu}} 
\institute{Department of Physics and Astronomy, Michigan State University,
  East Lansing,  MI 48824, USA, 
  \email{gilles@iaa.es}
  \and Instituto de Astrof\'isica de Andaluc\'\i a, 
  C/ Camino Bajo de Hu\'etor 50, 18008 Granada, Spain
  \and	GEPI/CAI, Observatoire de Paris, 77, Avenue Denfert-Rochereau,
  75014 Paris, France  
  \and    Departamento de F\'{i}sica, Universidad 
  de Concepci\'{o}n, Casilla 160-C, Concepci\'{o}n, Chile,
  \email{romanow@astro-udec.cl}
  \and    School of Physics and Astronomy, University of Nottingham, 
  University Park, Nottingham, NG7 2RD, UK
  \and    Department of Physics, University of Durham, South Road, Durham, 
  DH1 3LE, UK, 
  \email{r.m.sharples@durham.ac.uk}
  \and Department of Astronomy, Wesleyan University, Middletown, CT 06459, USA,
  \email{kathy@astro.wesleyan.edu}
  \and    Department of Astronomy, Yale University, P.\,O. Box 208101, 
  New Haven, CT 06520, USA
}
\date{Received DD Month YYYY / Accepted DD Month YYYY}
\abstract{
We present wide-field spectroscopy of globular clusters around the 
\object{Leo I group} galaxies \object{NGC~3379} and \object{NGC~3384} 
using the FLAMES multi-fibre instrument at the VLT.
We obtain  accurate radial velocities for 42 globular clusters (GCs) in 
total, 30 for GCs around the elliptical NGC~3379, eight around the 
lenticular NGC~3384, and four which may be associated with either galaxy.
These data are notable for their large radial range extending from
0\farcm7 to 14\farcm5 (2 to 42~kpc) from the centre of NGC~3379, and 
small velocity uncertainties of about 10~km\,s$^{-1}$.
We combine our sample of 30 radial velocities for globular
clusters around NGC~3379 with 8 additional GC 
velocities from the literature, and find a projected velocity
dispersion of $\sigma_{\rm p} = 175^{+24}_{-22}$~km\,s$^{-1}$ at 
$R < 5'$ and $\sigma_{\rm p} = 147^{+44}_{-39}$ at $R > 5'$. 
These velocity dispersions are consistent with a
dark matter halo around NGC~3379 with a concentration in
the range expected from a $\Lambda$CDM cosmological model
and a total mass of $\approx\,$6$\,\times\,10^{11} M_{\odot}$.
Such a model is also consistent with the stellar velocity dispersion 
at small radii and the rotation of the \ion{H}{i} ring at large radii,
and has a $(M/L)_\varB$ that increases by
a factor of five from several kpc to 100~kpc.
Our velocity dispersion  for the globular cluster system of NGC~3379 is 
somewhat higher than that found for the 
planetary nebulae (PNe) in the inner region covered by the PN data,
and we discuss possible reasons for this difference. 
 For NGC~3384, we find the GC system 
has a rotation signature broadly similar to that seen in other kinematic 
probes of this SB0 galaxy. This suggests that significant rotation
may not be unusual in the GC systems of disc galaxies.
}
\maketitle
\section{Introduction}

The current paradigm for galaxy formation is that galaxies form 
from baryons that cool within dark matter halos.
This paradigm is supported by the abundant observational evidence 
for dark matter in the Universe on many scales, ranging from clusters
of galaxies to the halos of individual spiral
and dwarf galaxies. Ideally, one would like to test for the 
presence of dark matter halos around early-type galaxies as well,
and to study the properties of these halos.

Determining the halo masses and mass profiles of individual 
elliptical galaxies has proven to be challenging, primarily because
they lack a readily observed dynamical tracer at large radii like the 
\ion{H}{i} gas of spiral galaxies.
Dedicated spectroscopic studies of the integrated light have been
limited to about $R \lesssim 2 R_{\rm e}$ (e.g., 
Kronawitter et al.\ \cite{kronawitter00}).
For the most luminous and massive ellipticals, 
X-ray observations of the hot gas around these galaxies
provide evidence for dark matter (e.g.\ Loewenstein \& 
White \cite{loewenstein99}). Studies of gravitational
lensing also indicate the presence of dark matter halos around
elliptical galaxies. Specifically, individual cases of strong 
lensing (e.g.,\ Treu \& Koopmans \cite{treu04}) show
evidence for dark matter at modest radii, while  
analyses of weak lensing (Wilson et al.\ \cite{wilson01}; 
McKay et al.\ \cite{mckay01}; 
Hoekstra et al.\ \cite{hoekstra04}) provide strong statistical evidence
for dark matter halos at large radii around early-type galaxies.  
Similar results for dark matter at large-scales
around elliptical galaxies are provided by studies of satellite
galaxies in large surveys (Prada et al.\ \cite{prada03};
Brainerd \cite{brainerd04}). However, for ordinary elliptical galaxies, 
a substantial range of profiles from inner to outer regions is allowed,
and the dependence of halo properties on galaxy luminosity is not strongly 
constrained. Therefore, determining the masses and mass profiles of ordinary 
individual elliptical galaxies is of significant interest.

The radial velocities of globular clusters (GCs) and planetary nebulae
around elliptical galaxies provide a powerful way to probe
the dynamics of their host galaxy. The extended spatial
distribution of GC systems makes them particularly
useful probes of any dark matter halos around elliptical 
galaxies. The value of these probes has been demonstrated
by the clear evidence for dark matter halos found by studies of
the radial velocities of GCs around the elliptical
galaxies NGC~4472 (e.g.\ Zepf et al.\ \cite{zepf00}; C\^ot\'e et al.\ 
\cite{cote03}), NGC~4486 (e.g., Cohen \cite{cohen00}; Romanowsky \&
Kochanek \cite{romanowsky01}),
NGC~1399 (Richtler et al.\ \cite{richtler04}), and NGC~5128
(Peng et al.\ \cite{peng04}). However, there are only
four galaxies with such high quality data, and with the exception of 
NGC~5128, all are considerably more luminous than a typical elliptical galaxy.

Spectroscopic studies of the planetary nebulae (PNe) can also probe the 
dynamics of elliptical galaxies. Romanowsky et al.\ (\cite{romanowsky03}),
hereafter R03, specifically carried out a study of the radial velocities of 
PNe around three elliptical galaxies of moderate luminosity. They found that 
the galaxy with the most data, NGC~3379, required 
little dark matter to the radial limit of their sample at
9~kpc ($3'$). The other two ellipticals, NGC~821 and NGC~4494
also appeared similar, but with weaker constraints. 
NGC~3379 ($\equiv$ \object{M105}) is in many ways the
archetypal ordinary elliptical galaxy, as it is the nearest
``normal'' giant elliptical galaxy, with a 
distance of 10 Mpc (Jensen et al.\ \cite{jensen03}; see also 
Gregg et al.\ \cite{gregg04}) 
in the Leo I group, and an intermediate luminosity of 
$L_\varB =$ 1.42 $\times 10^{10}L_{\odot}$ ($M_\varB = -19.9$).
Therefore, the possibility that its
dark matter halo might not be as expected is critical to follow up.

In this paper, we present the results of a study of the
dynamics of the GC system around NGC~3379. Studying the GC system has
two important advantages for understanding the halo dynamics
of NGC~3379 beyond the obvious utility of providing an independent 
test of the PNe results. Firstly, there has long been discussion
in the literature about the possibility that NGC~3379
has a significantly flattened component viewed face-on (Capaccioli et al.\ 
\cite{capaccioli91}). Subsequent papers have come to mixed conclusions  
about the presence of such a component (e.g., Statler \cite{statler94}; 
Statler \& Smecker-Hane \cite{statler99}; Statler \cite{statler01}), 
but it is obviously an important consideration for interpreting the radial 
velocities of objects observed in NGC~3379. 
Here, GCs provide a key advantage, because they are  
less likely than other tracer populations such as PNe to have
significant contribution from a strong disc component.  
The second advantage of GCs is that candidate GCs
can be identified in wide-field imaging of nearby 
ellipticals out to very large radii (e.g.\ Rhode \& Zepf \cite{rhode01},
\cite{rhode04} -- hereafter RZ04; Harris et al.\ \cite{harris04}). 
This large area over which GCs are observed allows the kinematics 
of the halo to be probed over a very large radial range.

The plan of the paper is that the spectroscopic observations 
and the determination of the radial velocities are presented
in Sect.~\ref{sect2}. The properties of the GC systems of the Leo I 
group galaxies NGC~3379 and NGC~3384 
are discussed in Sect.~\ref{sect3}. In Sect.~\ref{sect4}, we present the 
dynamical analysis of the NGC~3379 GC system and the implications for the dark 
matter halo of this galaxy. The conclusions are given in Sect.~\ref{sect5}.

\section{Observations and data analysis \label{sect2}}
\subsection{Identification of globular cluster candidates and MEDUSA 
fibre allocation}
The advent of large format images with mosaiced arrays of CCDs
on 4-m class telescopes allows for the identification of
candidate GCs over a wide field around galaxies
in the local universe. In particular, RZ04 carried out a
multi-colour imaging study of the GC systems of
nearby elliptical galaxies utilising the MOSAIC camera on the
{\it Mayall} 4 m telescope. They identified GC
candidates as unresolved objects with $\varB\varV\!\varR$  colours consistent
with those of Galactic GCs, taking into account
the photometric uncertainties (see RZ04 for more 
details). In order to obtain sufficient $S/N$ spectra in a 
$\la$\,4-hour exposure, we selected candidates brighter than 
$\varV \le 22$. To help minimise contamination from non-GCs,
we also placed a bright limit on the sample of 
$\varV \ge 19$, which affects only the extremely high luminosity tail of
the GC population at $M_\varV\,\la\,-11$. 
Over the $37' \times 37'$ MOSAIC field centered on NGC~3379, 
196 candidate GCs met these criteria for our spectroscopic follow-up. 

\begin{figure}
\resizebox{\hsize}{!}{\includegraphics{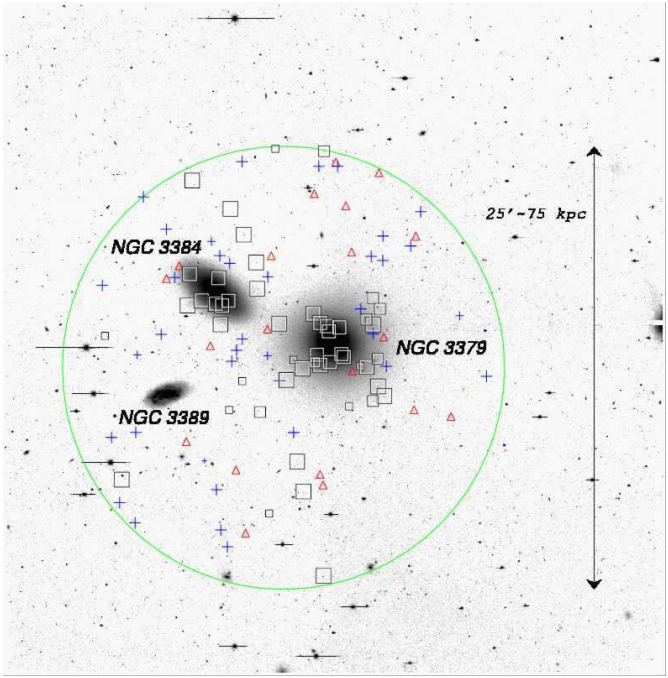}}
\caption{$\varV\!$-band MOSAIC image of the Leo I group. The image
covers $37' \times 37'$ and is plotted with north up and east to the 
left. The large circle shows the FLAMES 25$'$ diameter field of view,
corresponding to 75~kpc at the distance of NGC~3379.
The objects are noted with various symbols according to their
FLAMES radial velocities $\varv$ as follows: 
squares represent GCs ($500 < \varv < 1300$~km\,s$^{-1}$)
with larger squares indicating a more certain radial velocity,
\textcolor{blue}{crosses} represent objects identified as stars 
($\varv <$ 350~km\,s$^{-1}$) with the size of the
symbol corresponding to the confidence in the velocity as above, and
\textcolor{red}{triangles} show the 21 objects without a clear 
cross-correlation peak including objects with insufficient $S/N$
and background emission-line galaxies.}
\label{rv1}
\end{figure}

We used FLAMES, the multi-object, wide-field fibre spectrograph of 
VLT/UT2 (see Pasquini et al.\ \cite{pasquini04}) 
to efficiently obtain spectra of a significant 
sample of these GCs. The GIRAFFE/MEDUSA mode of this 
instrument offers 130 fibres over a $25'$ diameter field of
view, making it an excellent match to the wide-field MOSAIC images 
(see Fig.~\ref{rv1}) and the extended spatial distribution of the globular
cluster systems of nearby elliptical galaxies. 
We ran the FPOSS software to allocate the 130 available fibres 
to 108 GC candidates, twelve sky positions, and ten $\varV\!\sim18$ stars to 
be used as templates for radial velocity determinations, in a field centered at
$(\alpha,\delta)_{\mathit{J2000.0}} = (10^{\rm h}48^{\rm m}01^{\rm s}, 12\degr33'30'')$.

\subsection{Observations and radial velocity determination
\label{sect2.2}}
The FLAMES observations were carried out in service mode on VLT/UT2 
({\it Kueyen}), with the GIRAFFE/MEDUSA configuration.
We chose to use the grating LR4 which has a resolving power 
of $\lambda/\Delta(\lambda) \sim 6000$, and a central 
wavelength of $\lambda_{\rm c}=5431~$\AA\ that covers many absorption 
features over 800~\AA\ (including strong Mg and Fe lines), 
and sufficient efficiency to provide useful spectra ($S/N \ge 5$ per pixel) of 
$\varV \la 22$ unresolved sources in 3--4 hours. Our observing programme 
consisted of five 2595-s exposures taken on separate nights, 
in May--June 2003, under good seeing ($\sim$\,0\farcs7) conditions.

   \begin{figure}
   \resizebox{\hsize}{!}{\includegraphics[width=9cm]{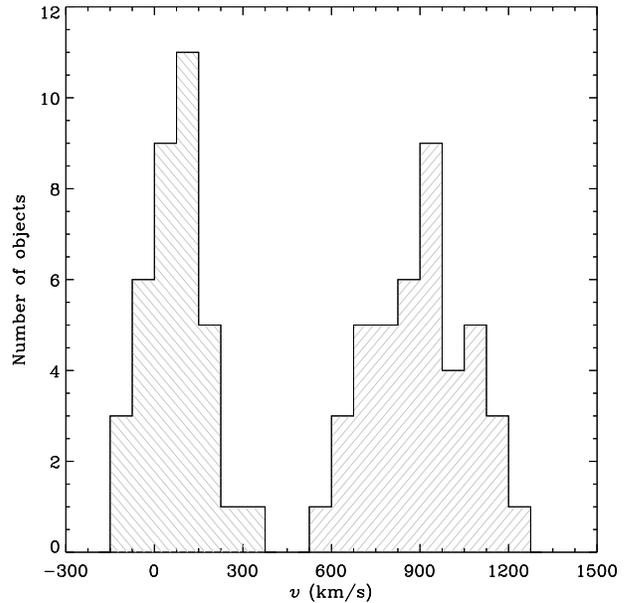}}
   \caption{
A histogram of the FLAMES radial velocities for all class A 
(34 GCs and 32 stars) and class B (8 GCs and 4 stars) objects. 
This histogram shows a clear $\sim$ 200~km\,s$^{-1}$ gap between the highest 
velocity stars and the lowest velocity clusters (associated with NGC~3384).}
   \label{histo}
   \end{figure}

After standard reductions to extract the spectra with the IRAF
{\textsc{hydra}} package, we determined the radial velocities $\varv$ of the 
objects by cross-correlating their spectrum with stellar templates. 
The cross-correlation function (CCF) peak was fitted 
by  the IRAF/{\sc rv} task {\tt fxcor} to estimate the velocity 
relative to the template, with a typical accuracy of 5--15~km\,s$^{-1}$ 
(each pixel represents 0.2 \AA\ or 11~km\,s$^{-1}$ with the LR4 grating). 

The cross-correlation was first done using the ten 
$\varV\!\sim$18 red stars to which we allocated fibres 
so as to provide stellar templates for radial velocity determination.
To supplement these ``simultaneous'' templates, we used 100 F5--K2 high 
signal-to-noise spectra -- both (sub)dwarfs and giants -- extracted from 
the \'ELODIE archive\footnote{{\texttt{http://atlas.obs-hp.fr/elodie}}} 
(Moultaka et al.\ \cite{moultaka04}). The \verb+LL_ELODIE+ 
stellar library is well-suited for GIRAFFE, with a very similar dispersion 
(the scale is also of 0.2~\AA/pixel, identical to the LR4 grating) which 
implies minimal rebinning of both spectra. 

The CCF peak Gaussian fits to all of the templates were checked 
both by determining the $\mathcal{R}_{_{\rm CCF}}$-value of the fit as 
defined by Tonry \& Davis (\cite{tonry79}), and additionally by eye. 
These agreed well, and the velocities obtained for
a dozen of the brightest targets were compared between individual exposure
spectra, and also showed good agreement to within the errors estimated 
by {\tt fxcor}. We then adopted the following classification of the quality of
FLAMES radial velocity determinations: 
\begin{itemize}
\item class A: more than 98\% of \'ELODIE and more than five (out of ten)
simultaneous templates agree within the errors $\delta\varv$; 
\item class B: more than 80\% of \'ELODIE and $\ge$\,3 simultaneous 
templates agree within the estimated errors;
\item class C {\it (not used)}: less than 60\% of \'ELODIE and only
0--2 simultaneous templates agree.
\end{itemize}
 For one object, \textsf{gc771}, we clearly detected 
[\ion{O}{iii}]\,$\lambda$\,5007 at a radial velocity of 766~km\,s$^{-1}$.
This both provides interesting evidence for a planetary nebula
in an extragalactic GC, suggesting these may not be exceedingly rare 
(see Minniti \& Rejkuba \cite{minniti02} for another example),
and also confirmation of the best estimate for the radial
velocity from the cross-correlation of the absorption lines.
For all subsequent work, we consider the objects with class A and B 
radial velocities, and defer consideration
of the class C objects until further data are obtained. 

The reliability and accuracy of our radial velocities are strongly
supported by a comparison with the data of Puzia et al.\ \cite{puzia04} 
(hereafter P04) who used the FORS instrument on the VLT to obtain radial
velocities for 18 NGC~3379 inner GCs. Ten of the clusters in the P04 dataset 
are in our sample, including nine of our class A objects and one class B
cluster. 
The agreement between the two studies is excellent with an average 
difference of 1~km\,s$^{-1}$ and a dispersion of 30~km\,s$^{-1}\!$. 
Since 30~km\,s$^{-1}$ is close to the typical uncertainty in the lower
spectral resolution P04 data, this comparison is 
consistent with the small uncertainties we find for our
velocities of only about 10~km\,s$^{-1}$ in most cases. 

\subsection{Separation of globular clusters and stars}
The histogram of measured radial velocities for the objects we observed
is shown in Fig.~\ref{histo}. The 
heliocentric velocities of NGC~3379 and NGC~3384 are
904~km\,s$^{-1}$ and 721~km\,s$^{-1}$ respectively 
(Falco et al.\ \cite{falco99}; Smith et al.\ \cite{smith00}) and the
galaxies are expected to have modest velocity dispersions.
As a result, GCs are well separated from stars in the data: there is a 200 
km\,s$^{-1}$ gap between the lowest velocity object we consider a globular 
cluster and the highest velocity star. Specifically, we classify objects with 
500 $< \varv <$ 1300~km\,s$^{-1}$ as GCs, and objects with
$-102\le \varv \le332$~km\,s$^{-1}$ as stars. The latter range is in good 
agreement with the expected star field velocity of about 60~km\,s$^{-1}$
in the direction of NGC~3379 at $(l,b)=(233\fdg5,57\fdg6)$. 

We then have a total sample of 42 GCs with radial 
velocities, listed in Table~\ref{vit} along with various properties; 
34 of these GCs have highly confident class A radial velocities, and 8
have likely class B determinations. These form the basis for the
analysis and discussion in subsequent sections.

We also have radial velocities for 36 stars in our field. This is consistent
with expectations from the photometric survey of RZ04,
which predicts that about 60\% of our objects at a distance of 4$'$ 
from NGC~3379 are GCs, in agreement with the 22 GCs and eleven non-GCs
(stars and unclassified objects) within this radius.
 Most unclassified targets have $\varV\ga 21$ and are probable galaxies, 
compact but unresolved in the MOSAIC images (typical seeing 1\farcs3) 
-- including  [\ion{O}{ii}] emission line objects 
at 0.35\,$\la$\,$z$\,$\la$\,0.55 for the LR4 wavelength range. 
The total number of stars and background objects found is consistent 
with the~work of RZ04, given our fibre allocation and $\varV\!\le$~22~limit.

\section{Properties of the globular cluster systems of NGC~3379 and NGC~3384
\label{sect3}}

\begin{table*}
\caption{FLAMES heliocentic radial velocities $\varv$ of globular clusters 
in Leo I: for each of the 42 confirmed GCs,
$\varv$ and its error estimated by {\tt fxcor} are the median 
values over the \'ELODIE templates (ordered by projected distance to 
NGC~3379, $R$ in arcsec; the position 
angle $\theta$ is counted in degrees from 0 to 90, N to E). 
Numbering is from Rhode \& Zepf (\cite{rhode04}) 
 as well as the $\varV$ magnitudes, $\varB-\varV$ and $\varV-\varR$ colours. 
Equatorial {\it J2000.0} coordinates should be accurate to 
$\lesssim$\,0\farcs2 with respect to the GSC2.2. 
 The Tonry \& Davis (\cite{tonry79}) cross-correlation coefficient  
$\mathcal{R}_{_{\rm CCF}}$ is the median value over the \'ELODIE templates. 
Class B objects are commented, as well as the 10 GCs in common with 
Puzia et al.\ (\cite{puzia04}) with their velocity estimate. 
Full tables of all MEDUSA targets -- including stars and 
galaxies/unclassified objects -- are available online. 
} 
\label{vit}
\begin{tabular}{l c c r l c c c r rcl l}
\hline\hline
Id$_{\rm{RZ04}}$&$\alpha~$({\it J2000.0}) & $\delta\ $({\it J2000.0})&$R~\arcsec$&$\theta~\degr$&$\varV$&$\varB$$\!-\!\varV$&$\varV\!\!-$$\!\varR$&$\mathcal{R}_{_{\rm CCF}}$&\multicolumn{3}{c}{$\varv\pm\delta\varv$~~(km\,s$^{-1}$)} & Comments\\
\hline
{\textsf{gc771}}
                 &10$^{\rm h}$47$^{\rm m}$50\fs62&12\degr35\arcmin31\farcs9&  39 & 20 &21.26  &0.68&0.52 &6.0 & $758$&$\pm$&$9 $&Puzia23 at 791~km\,s$^{-1}$\\ 
{\textsf{gc830}} &10$^{\rm h}$47$^{\rm m}$47\fs71&12\degr34\arcmin14\farcs5&  49 &217 &21.11  &0.85&0.59 &16.4& $926$&$\pm$&$5 $\\
{\textsf{gc820}} &10$^{\rm h}$47$^{\rm m}$48\fs18&12\degr35\arcmin44\farcs9&  55 &336 &20.94  &0.75&0.49 &9.9 & $889$&$\pm$&$6 $&Puzia27 at 911~km\,s$^{-1}$\\
{\textsf{gc764}} &10$^{\rm h}$47$^{\rm m}$51\fs07&12\degr35\arcmin49\farcs1&  58 & 21 &20.00  &0.72&0.52 &18.7& $776$&$\pm$&$4 $&Puzia24 at 801~km\,s$^{-1}$\\
{\textsf{gc839}} &10$^{\rm h}$47$^{\rm m}$47\fs19&12\degr34\arcmin04\farcs6&  62 &217 &21.98  &0.69&0.46 &6.7 & $1064$&$\pm$&$13$\\
{\textsf{gc774}} &10$^{\rm h}$47$^{\rm m}$50\fs40&12\degr33\arcmin47\farcs7&  67 &171 &21.36  &0.68&0.50 &5.3 & $656$&$\pm$&$9 $&Puzia14 at 645~km\,s$^{-1}$\\
{\textsf{gc728}} &10$^{\rm h}$47$^{\rm m}$53\fs42&12\degr34\arcmin12\farcs2&  69 &127 &21.83  &0.81&0.44 &5.2 & $978$&$\pm$&$15$&Puzia13 at 941~km\,s$^{-1}$\\
{\textsf{gc741}} &10$^{\rm h}$47$^{\rm m}$52\fs60&12\degr35\arcmin59\farcs6&  78 & 34 &21.27  &0.97&0.59 &9.6 & $1188$&$\pm$&$6$\\
{\textsf{gc740}} &10$^{\rm h}$47$^{\rm m}$52\fs65&12\degr33\arcmin37\farcs9&  88 &150 &21.85  &1.07&0.52 &6.2 & $889$&$\pm$&$7 $&Puzia12 at 867~km\,s$^{-1}$\\
{\textsf{gc719}} &10$^{\rm h}$47$^{\rm m}$53\fs83&12\degr33\arcmin46\farcs2&  92 &138 &21.51  &0.65&0.52 &4.4 & $740$&$\pm$&$10$&Puzia11 at 705~km\,s$^{-1}$; class B\\
{\textsf{gc709}} &10$^{\rm h}$47$^{\rm m}$54\fs19&12\degr36\arcmin31\farcs5& 118 & 35 &20.97  &0.72&0.50 &13.0& $954$&$\pm$&$4 $\\
{\textsf{gc682}} &10$^{\rm h}$47$^{\rm m}$56\fs72&12\degr33\arcmin26\farcs4& 137 &130 &20.47  &0.63&0.48 &8.1 & $1137$&$\pm$&$6 $&Puzia09 at 1130~km\,s$^{-1}$\\
{\textsf{gc923}} &10$^{\rm h}$47$^{\rm m}$41\fs91&12\degr36\arcmin11\farcs2& 139 &303 &21.97  &0.75&0.57 &4.0 & $999$&$\pm$&$14$ & Class B\\
{\textsf{gc925}} &10$^{\rm h}$47$^{\rm m}$41\fs87&12\degr33\arcmin31\farcs0& 144 &235 &21.18  &0.99&0.56 &13.2& $1048$&$\pm$&$4$\\
{\textsf{gc948}} &10$^{\rm h}$47$^{\rm m}$40\fs34&12\degr35\arcmin54\farcs9& 152 &293 &20.25  &0.66&0.38 &6.8 & $1264$&$\pm$&$9$&Puzia33 at 1310~km\,s$^{-1}$\\
{\textsf{gc958}} &10$^{\rm h}$47$^{\rm m}$39\fs34&12\degr33\arcmin59\farcs1& 164 &250 &21.57  &0.74&0.42 &4.4 & $951$&$\pm$&$8 $ & Class B\\
{\textsf{gc634}} &10$^{\rm h}$48$^{\rm m}$02\fs13&12\degr35\arcmin57\farcs3& 196 & 71 &20.97  &0.77&0.46 &7.0 & $950$&$\pm$&$6 $\\
{\textsf{gc971}} &10$^{\rm h}$47$^{\rm m}$38\fs75&12\degr36\arcmin47\farcs8& 199 &305 &21.78  &0.83&0.47 &4.4 & $1027$&$\pm$&$13$ & Class B\\
{\textsf{gc949}} &10$^{\rm h}$47$^{\rm m}$40\fs33&12\degr37\arcmin24\farcs1& 205 &317 &21.81  &0.69&0.43 &3.0 & $933$&$\pm$&$27$ & Class B\\
{\textsf{gc645}} &10$^{\rm h}$48$^{\rm m}$00\fs52&12\degr32\arcmin46\farcs9& 206 &128 &21.02  &0.78&0.48 &7.5 & $654$&$\pm$&$6 $&Puzia05 at 632~km\,s$^{-1}$\\
{\textsf{gc960}} &10$^{\rm h}$47$^{\rm m}$39\fs25&12\degr32\arcmin25\farcs6& 216 &226 &20.06  &0.81&0.49 &14.9& $899$&$\pm$&$5 $\\
{\textsf{gc946}} &10$^{\rm h}$47$^{\rm m}$40\fs46&12\degr31\arcmin36\farcs4& 241 &215 &21.67  &0.65&0.36 &4.3 & $739$&$\pm$&$6 $ & Class B\\
{\textsf{gc984}} &10$^{\rm h}$47$^{\rm m}$37\fs80&12\degr31\arcmin52\farcs7& 254 &224 &20.86  &0.82&0.49 &7.8 & $1130$&$\pm$&$5$\\
{\textsf{gc573}} &10$^{\rm h}$48$^{\rm m}$07\fs19&12\degr37\arcmin55\farcs1& 318 & 55 &20.94  &0.95&0.57 &13.0& $752$&$\pm$&$4 $&NGC~3384 cluster?\\
{\textsf{gc581}} &10$^{\rm h}$48$^{\rm m}$06\fs60&12\degr30\arcmin58\farcs5& 346 &133 &21.82  &0.98&0.58 &3.9 & $1064$&$\pm$&$11$ & Class B\\
{\textsf{gc571}} &10$^{\rm h}$48$^{\rm m}$07\fs40&12\degr39\arcmin24\farcs1& 378 & 45 &21.25  &0.73&0.50 &5.9 & $792$&$\pm$&$9 $&NGC~3384 cluster?\\
{\textsf{gc482}}$^{\rm a}$&10$^{\rm h}$48$^{\rm m}$13\fs88&12\degr37\arcmin14\farcs1& 388 & 69 &21.59&1.00&0.63 &17.6& $908$&$\pm$&$6 $&NGC~3384 cluster\\
{\textsf{gc454}} &10$^{\rm h}$48$^{\rm m}$15\fs73&12\degr35\arcmin54\farcs2& 394 & 81 &21.89&1.06&0.62 &6.1 & $874$&$\pm$&$8 $&NGC~3384 cluster\\
{\textsf{gc461}} &10$^{\rm h}$48$^{\rm m}$15\fs36&12\degr36\arcmin58\farcs0& 404 & 72 &21.87&0.71&0.45 &9.4 & $847$&$\pm$&$17$&NGC~3384 cluster\\
{\textsf{gc670}} &10$^{\rm h}$47$^{\rm m}$58\fs08&12\degr28\arcmin11\farcs2& 422 &163 &20.37  &0.72&0.44 &7.4 & $921$&$\pm$&$6 $\\
{\textsf{gc442}} &10$^{\rm h}$48$^{\rm m}$16\fs81&12\degr37\arcmin04\farcs0& 426 & 72 &21.15&0.74&0.51 &7.7 & $878$&$\pm$&$11$&NGC~3384 cluster\\
{\textsf{gc449}} &10$^{\rm h}$48$^{\rm m}$16\fs20&12\degr38\arcmin30\farcs6& 452 & 61 &20.97&0.80&0.54 &15.8& $672$&$\pm$&$4 $&NGC~3384 cluster\\
{\textsf{gc387}} &10$^{\rm h}$48$^{\rm m}$20\fs06&12\degr37\arcmin15\farcs5& 476 & 73 &21.98&0.83&0.44 &5.4 & $704$&$\pm$&$11$&NGC~3384 cluster\\
{\textsf{ad1102}}&10$^{\rm h}$48$^{\rm m}$10\fs29&12\degr40\arcmin57\farcs9& 476 & 40 &19.28  &0.75&0.50 &23.8& $920$&$\pm$&$3 $ &
NGC~3384 cluster? \\
{\textsf{gc683}} &10$^{\rm h}$47$^{\rm m}$56\fs61&12\degr26\arcmin30\farcs2& 514 &168 &21.14  &0.70&0.39 &4.4 & $1087$&$\pm$&$14$\\
{\textsf{gc351}} &10$^{\rm h}$48$^{\rm m}$23\fs34&12\degr36\arcmin57\farcs7& 519 & 76 &21.60&1.02&0.49 &3.9 & $739$&$\pm$&$17$&NGC~3384 cluster\\
{\textsf{gc358}} &10$^{\rm h}$48$^{\rm m}$22\fs79&12\degr38\arcmin44\farcs2& 546 & 65 &21.35&0.59&0.46 &5.6 & $577$&$\pm$&$13$&NGC~3384 cluster\\
{\textsf{ad1021}}&10$^{\rm h}$48$^{\rm m}$13\fs33&12\degr42\arcmin24\farcs8& 573 & 38 &21.71  &0.99&0.52 &5.3 & $752$&$\pm$&$7 $&NGC~3384 cluster?\\
{\textsf{gc756}} &10$^{\rm h}$47$^{\rm m}$51\fs58&12\degr45\arcmin40\farcs2& 646 &  2 &20.87  &0.58&0.47 &3.7 & $725$&$\pm$&$19$ & Class B\\
{\textsf{gc368}} &10$^{\rm h}$48$^{\rm m}$22\fs19&12\degr44\arcmin00\farcs3& 731 & 42 &21.15  &0.82&0.44 &4.6 & $1098$&$\pm$&$10$& \\
{\textsf{ad1481}}&10$^{\rm h}$47$^{\rm m}$52\fs08&12\degr21\arcmin45\farcs2& 790 &177 &19.39  &0.52&0.43 &7.0 & $1105$&$\pm$&$6$\\
{\textsf{gc185}} &10$^{\rm h}$48$^{\rm m}$38\fs75&12\degr27\arcmin09\farcs3& 870 &122 &21.09  &0.64&0.37 &4.8 & $964$&$\pm$&$10$\\
\hline
\end{tabular}
$^{\mathrm{a}}$ $\equiv$ Brodie \& Larsen (\cite{brodie02})  
N3384--FF--7 ``fuzzy'' cluster. They found $\varv = 905\pm 23$ 
km\,s$^{-1}$  (red channel value) using {\it Keck}\,/\,LRIS.  
\end{table*}

\subsection{Identifying globular  
clusters with NGC~3379 and NGC~3384 \label{sect3.2}}

The core of the Leo I group has a compact appearance on the sky with the 
central elliptical (E1) galaxy NGC~3379 separated by only $7'$ from the nearby
SB0 galaxy NGC~3384, and by $10'$ to the small spiral NGC~3389,
corresponding to projected distances of 20 and 29~kpc respectively.
However, the galaxies are fairly clearly delineated in their
radial velocities, with NGC~3379 at 904~km\,s$^{-1}$,
NGC~3384 at 721~km\,s$^{-1}$, and NGC~3389 with 1300~km\,s$^{-1}$.
The velocity for NGC~3379 is 
 the average of the Updated Zwicky Catalogue (UZC, Falco 
 et al.\ \cite{falco99}) and Smith et al.\ (\cite{smith00}) values, 
 and the velocities for NGC~3384 and NGC~3389 are from the UZC.
 The scatter of measurements in the literature suggests uncertainties 
 of 10--20~km\,s$^{-1}$ in these numbers. 

\begin{figure}
\resizebox{\hsize}{!}{\fbox{\includegraphics[angle=0]{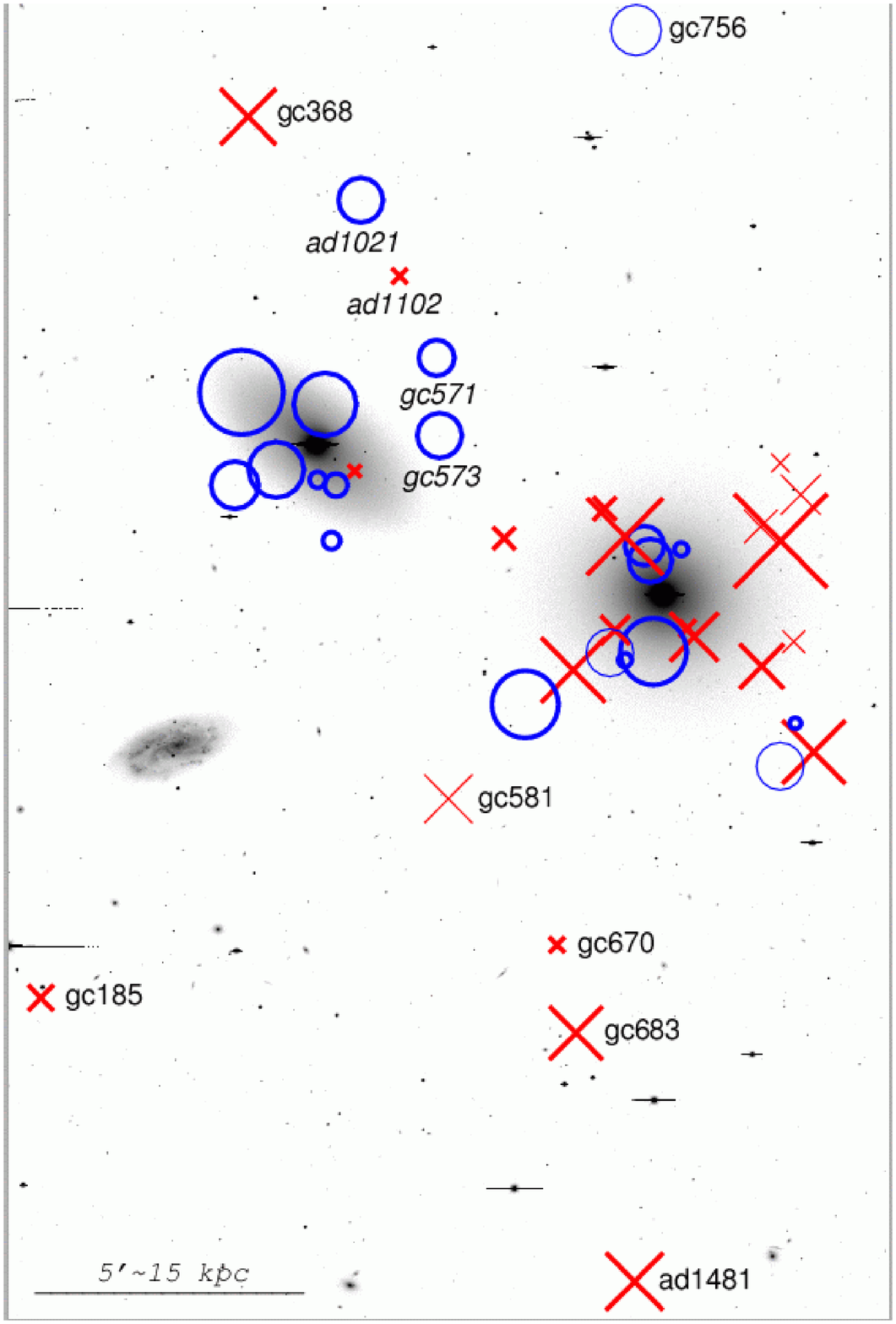}}}
\caption{Velocities of the 42 confirmed GCs overplotted on~a 
$\varV\!$-band MOSAIC sub-image of $25' \times 15'$ 
(north is up, east to the left). Crosses (\textcolor{red}{$\times$}) 
are GCs with $\varv >$ 904~km\,s$^{-1}$ which we adopt as the 
systemic velocity of NGC~3379. Circles (\textcolor{blue}{$\bigcirc$}) are 
those GCs with $\varv <$ 904~km\,s$^{-1}$; the symbol size is 
proportional to the offset, up to about $\pm$400~km\,s$^{-1}$, with class A
objects in boldface. Note the clear rotation of the eight GCs along the 
NGC~3384 disc and their lower typical velocities compatible with the 
systemic value of this SB0. The seven outer ($R>5'$) GCs
associated with NGC~3379 and the four globulars with uncertain
assignments to one of the two galaxies are identified individually,
with the unassigned objects in italics.
}
\label{rv2}
\end{figure}

In Fig.~\ref{rv2} we show the location and  
velocities of our GC sample overlayed on an image of the Leo I 
group. This figure shows the clear velocity offset of the
group of GCs around NGC~3379, which all scatter
around the galaxy's systemic velocity, from the group around 
NGC~3384, which have velocities like that of
NGC~3384 and are thus offset to lower $\varv$ 
than NGC~3379 and its GC system. The velocity difference
of these two galaxies and their GC systems is 
useful for determining which clusters in their outer halos
belong to which galaxies given their closeness in projection on the sky. 

Based on their radial velocities and positions, we assign all of the GCs 
spatially centered around NGC~3379 to its GC system, along with the GCs 
located in intermediate spatial regions that have velocities greater than 
the systemic velocity of NGC~3379, and the globular {\sf gc756}. 
We also assign the eight GCs centered on NGC~3384 to this galaxy, and
note that these have a strong rotation signature, which
we discuss further in Sect.~\ref{sect3.3}. We consider the objects 
\textsf{\sf{gc573}}, \textsf{\sf{gc571}}, 
\textsf{\sf{ad1102}}, and \textsf{\sf{ad1021}} as possibly
belonging to either galaxy. We note that it would seem unlikely
that NGC~3379 has only outer globulars to its south and not
its north, so some of these four would seem likely to belong to
NGC~3379. On the other hand they are slightly closer in projection 
to NGC~3384, so some of the four would appear likely to be associated
with its halo. The velocities of these four GCs are also generally 
consistent with NGC~3384 (see Fig.~\ref{mydisp}), but if none of them 
is assigned to NGC~3379, then its outer halo would have a strange 
bias towards only globulars with velocities greater than both 
the systemic velocity of NGC~3379 and its inner GC system (a similar argument 
supports the identification of {\sf gc756} with NGC~3379). 
We also note that although at very 
large radii it may become more sensible to consider the halo
of the group of galaxies together, given the clear velocity
offset of the galaxies and their inner GC
systems, a closer relationship to the individual galaxies
is indicated over our field of view of $\approx$ 30~kpc. 

For the calculations that follow, we use the assignments
described above and given in Table~\ref{vit}, and also
consider the effect on these calculations of including
or excluding various combinations of the GCs
with uncertain assignments. As noted below, we find that
the conclusions are not significantly affected by the
choices in these assignments.

\subsection{Kinematics of the globular cluster system of NGC~3379}
Our sample of NGC~3379 GCs with radial velocities consists of the 30 clusters 
clearly identified with NGC~3379 and the four for which this assignment 
is uncertain. To these, we add eight additional GCs from the study of P04 
(see Sect.~\ref{sect2.2}). 

\begin{figure}
   \resizebox{\hsize}{!}{\includegraphics[width=9cm]{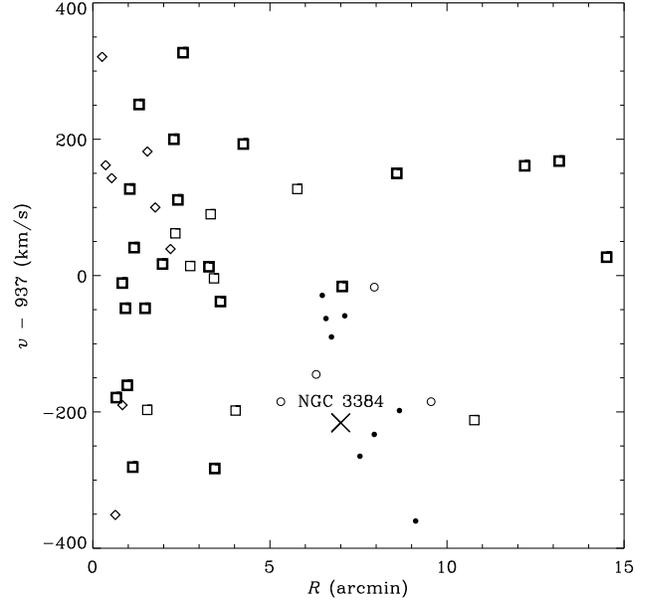}}
      \caption{Relative velocities vs.\ distances to NGC~3379 for the 
spectroscopic sample of 42 globular clusters. GCs associated with NGC~3379 
are shown as squares (bold for class A; diamonds are eight additional 
GCs from P04), those associated with NGC~3384 as filled circles
and GCs possibly associated with either galaxy as open circles.}
   \label{mydisp}
   \end{figure}

For the sample of 38 GCs associated with NGC~3379, we find a mean of
$\varv = 947$~km\,s$^{-1}$ with an uncertainty of 38~km\,s$^{-1}$ determined
via Monte Carlo simulations. The projected velocity dispersion of the
full sample is $\sigma_{\rm p} = 169^{+20}_{-20}$~km\,s$^{-1}$. 
We discuss the $\sigma_{\rm p}$ as a function of radius
in the following section, in which we use this to constrain 
the mass of the halo of NGC~3379. One additional note
is that the mean value of the GC radial velocities is 
$1\sigma$ larger than the velocity of the galaxy at 
904~km\,s$^{-1}$ (Sect.~\ref{sect3.2}). 
 Our GIRAFFE data have good spectral resolution and are 
highly repeatable from night to night.
Moreover, there is excellent  agreement between our radial 
velocities and the independent data of P04. 
Thus, the most likely explanation is that the small offset
is a statistical fluctuation, which can be tested with more data.
For modelling purposes, we will use hereafter
the mean GC velocity of 937~km\,s$^{-1}$ 
from the central regions with good azimuthal coverage (Fig.~\ref{mydisp}).

Our data also show little evidence for rotation in the NGC~3379
GC system. Although our sample size is modest, 
even with a much smaller sample we are able to detect rotation in 
NGC~3384 owing to our small velocity uncertainties (see Sect.~\ref{sect3.3}).
The exact constraint on the rotation around NGC~3379 depends 
somewhat on the sample definition, but by comparison 
to the strong signal seen around NGC~3384, the overall rotation in
the NGC~3379 GC system appears to be 
less than 100~km\,s$^{-1}$ about any axis. This low rotation is consistent 
with that seen in the stars at small radii (e.g.\ Statler \& Smecker-Hane 
\cite{statler99}) and is similar to that seen in the GC system of the 
elliptical galaxy NGC~4472 (see Zepf et al.\ \cite{zepf00}; 
C\^ot\'e et al.\ \cite{cote03}) although probably less than that in NGC~5128 
(Peng et al.\ \cite{peng04}), a peculiar merger remnant.

\subsection{Kinematics of the globular cluster system of NGC~3384 
\label{sect3.3}}
Our sample also has eight GCs that
are clearly associated with NGC~3384. Even with just these
eight radial velocities, there is  
 evidence for rotation ($v / \sigma \sim$ 2) in the GC system of 
NGC~3384, which can be seen visually in Fig.~\ref{rv2}. Specifically, 
the four GCs to the south-west of the galaxy 
centre have radial velocities 200~km\,s$^{-1}$ lower than the 
four to the north-east. This rotation signature is very 
similar to that found by a study of 68 PNe 
around this galaxy by Tremblay et al.\ (\cite{tremblay95}).
The rotation in the GC system of NGC~3384 suggests
that rotation may not be unusual in the GC systems
of disc galaxies. This brings into question whether the lack
of rotation in the large majority of the Milky Way globular
cluster system (e.g.\ Zinn \cite{zinn85}) is somewhat unusual, and the 
rotation seen in the Messier 31 GC system (e.g., Perrett
et al.\ \cite{perrett02}) is possibly more typical. 

\subsection{Colours of the globular clusters around NGC~3379 and NGC~3384}
\begin{figure}
   \resizebox{\hsize}{!}{\includegraphics[width=9cm]{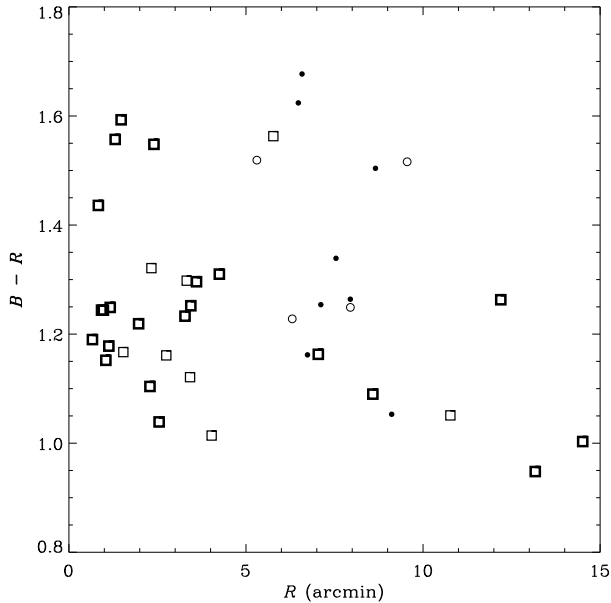}}
    \caption{$\varB\!-\!\varR$ colours vs.\ radii for our spectroscopic sample 
of 42 GCs (symbols are as in Fig.~\ref{mydisp}). 
The plot shows the trend of redder colours  for GCs close to the NGC~3379 
centre and bluer colours at large radii.}
\label{b-r}
   \end{figure}

We can use our dataset to examine questions about the colours
of the NGC~3379 and NGC~3384 GCs with a spectroscopically 
confirmed dataset. In Fig.~\ref{b-r} we plot the $\varB-R$ colours
from RZ04 of our GC sample as a function of their
distance from the centre of NGC~3379. The data show that NGC~3379
has more  blue, metal-poor GCs than red metal-rich
ones, in agreement with the suggestion of RZ04 from their
photometry. As noted in that paper, of current models for
the formation of GC systems, only those that 
involve galaxy mergers (e.g.\ Ashman \& Zepf \cite{ashman92}; 
Beasley et al.\ \cite{beasley02}) are readily able to account for this 
larger number of blue relative to red GCs in an elliptical galaxy like 
NGC~3379 with its modest luminosity and small GC population.

The data shown in Fig.~\ref{b-r} also indicate a colour
gradient in the GC system of NGC~3379, such
that the ratio of blue to red GCs increases
with distance away from the galaxy centre. This is consistent
with RZ04 and many other previous studies for the globular
cluster systems of a number of elliptical galaxies (see 
Ashman \& Zepf \cite{ashman98} for a review). However, RZ04
also noted that the photometry indicated that there may be red,
metal-rich GCs at large distances around
some elliptical galaxies. In NGC~3379, our data show a 
 metal-rich GC ({\sf gc581}) at a projected distance of almost $6'$, 
corresponding to about 17~kpc.  The GC {\sf ad1021} 
is also interesting in this regard, as it is
very red, and is  almost $10'$ away from NGC~3379 and 
about $5'$ away from NGC~3384. Thus there are two spectroscopically confirmed, 
metal-rich clusters well into the halos of these galaxies. 
 We also note that two of the inner GCs
associated with NGC~3384 are very red, which might be
attributable to  dust in the system. 

\section{Constraints on the halo of NGC~3379 \label{sect4}}

\subsection{Velocity dispersion profile}
\begin{figure}
   \resizebox{\hsize}{!}{\includegraphics[width=9cm]{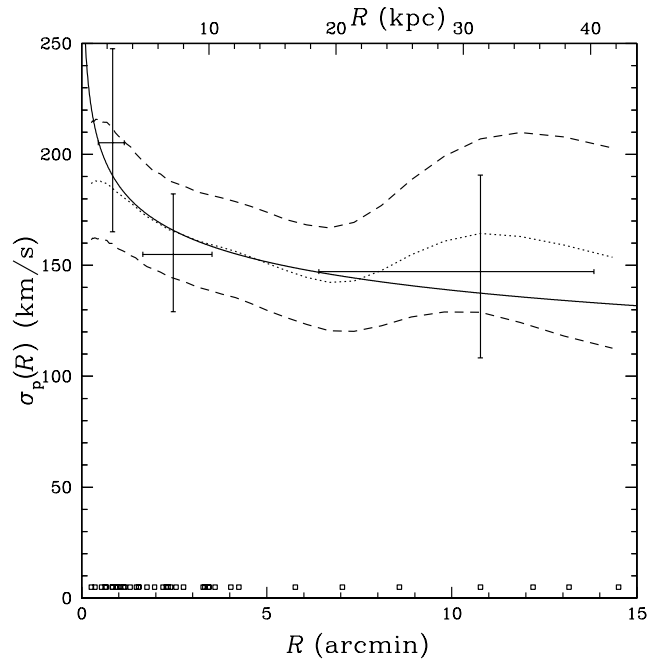}}
      \caption{Projected velocity dispersion vs. radius for the NGC~3379 GCs. 
The error bars show radially-binned data, where the vertical bars 
are for the $\pm 34\%$ uncertainties, and the horizontal bars show the range
covered by 68\% of the data.
The dotted line shows a moving window estimate, with dashed lines signifying
the $\pm 34\%$ uncertainties. The solid line shows the best-fit scale-free 
power-law profile. The boxes at the bottom indicate the radial positions 
of the GC velocity data.}
         \label{disp}
   \end{figure}

One of the primary goals of this study is to determine the radial 
velocities of GCs to probe the halo dynamics of NGC~3379.
To test mass models of the halo of NGC~3379, we use our dataset
to determine the projected velocity dispersion at several different 
radii. As shown in Fig.~\ref{disp}, we adopt three radial bins
to allow the comparison of the mass over a range of radii,
corresponding in the inner region to the largest radial
extent of the stellar data, in the intermediate region 
to the largest extent of the PNe data, and
the outer region which is uniquely probed by our globular
cluster dataset. We then plot the projected velocity 
dispersions $\sigma_{\rm p}$ in these three radial regions in 
Fig.~\ref{disp}.  To calculate $\sigma_{\rm p}$, the 
mean velocity of $\varv_0=937$~km\,s$^{-1}$ is adopted, 
and the measurement uncertainties $\delta \varv_i$ are taken into account.
The $\pm 34\%$ uncertainties are estimated by 30\,000 Monte Carlo 
simulated data sets drawn from the best-fit model.
Note that the dispersion $\sigma_{\rm p}$ discussed throughout this paper 
incorporates any rotational support, and thus is technically 
$\varv_{\rm rms}=(\varv_{\rm p}^2 + \sigma_{\rm p}^2)^{1/2}$.

Figure~\ref{disp} shows the binned velocity dispersion profile. We estimate
$\sigma_{\rm p} = 205^{+43}_{-40}$~km\,s$^{-1}$, $155^{+27}_{-26}$~km\,s$^{-1}$
and $147^{+44}_{-39}$~km\,s$^{-1}$ at $R$$\sim$0\farcm8 (13 GCs), 
$\sim$2\farcm5 (18 GCs) and $\sim$11' (7 GCs), respectively.
Inclusion of the four uncertain GCs has very little effect: the outer 
$\sigma_{\rm p}$ becomes $147^{+33}_{-32}$~km\,s$^{-1}$. Alternatively 
adopting $904$~km\,s$^{-1}$ for the NGC~3379 velocity  
 has the effect of raising the outer two $\sigma_{\rm p}$ estimates 
 to $165^{+29}_{-27}$ km\,s$^{-1}$ and $165^{+48}_{-44}$ km\,s$^{-1}$.
 
 We also construct a smoothed estimate of $\sigma_{\rm p}(R)$, using a 
``moving window''. This performs a $\sigma_{\rm p}$ estimate at every radius 
$R$ using the nearby data (each estimate is thus not independent). Unlike 
previous versions of this method which used a window of constant width (Zepf 
et al.\ \cite{zepf00}; C\^{o}t\'{e} et al.\ \cite{cote01}) or with fixed 
number of data points (Kissler-Patig \& Gebhardt \cite{kissler98}; 
Cohen \cite{cohen00}), we use an exponential weighting function for the 
contribution of each data point $\varv_i$ to the estimate at radius $R$:
\begin{equation}
w_i(R)=\frac{1}{\sigma_R} \exp\left[\frac{(R-R_i)^2}{2\sigma_R^2}\right] ,
\end{equation}
where the half-width $\sigma_R \propto \left[R \Sigma(R)\right]^{-1}$,
and $\Sigma(R)$ is the underlying surface density distribution from
which the measurements are drawn. For ``perfect'' sampling, the number 
of points contributing to each estimate is constant.
For optimal smoothing, the number of points within $\pm \sigma_R$
has a maximum of $\sim$\,25.
For the current data set, the surface density is the $R^{1/4}$ law found by 
RZ04, and the window width is normalised to $\sigma_R \simeq 1\farcm8$.

Figure~\ref{disp} shows the result of this procedure.
The dispersion profile is consistent with being constant over
the wide range of radii studied, with a possible increase to small radii.
 The inclusion of the four uncertain GCs does not affect the profile much,
but a mean velocity of $904$~km\,s$^{-1}$ would raise $\sigma_{\rm p}$ in
most places by 10--20~km\,s$^{-1}$. Utilising a maximum likelihood
technique for fitting the data to a scale-free power-law (see Appendix A),
we also find the same result that the velocity dispersion is generally
constant with radius with a possible modest increase in the dispersion 
to small radii.

\subsection{Halo models}

A primary goal of this programme is to test whether our measured
velocity dispersion profile for the GC system of NGC~3379 is consistent 
with the mass distribution expected from standard $\Lambda$CDM 
halos or other models. To perform this test, we construct mass
models of NGC~3379 with and without dark matter halos.
The circular velocity of these models is:
\begin{equation}
\varv^2_{\rm c}(r) = \frac{\varv_\star^2 a_\star r}{(r+a_\star)^2} +
\varv_s^2\left[\frac{r_s}{r}\ln\left(1+\frac{r}{r_s}\right)-\frac{r_s}{r+r_s}\right] ,
\label{eq2}
\end{equation}
where $\varv_{\star}$ and $a_{\star}$ describe the stellar mass
distribution (after Hernquist \cite{hernquist90}) 
 and $\varv_s$ and $r_s$ are the Navarro, Frenk \& White (\cite{navarro96})
 NFW parameters of the dark matter halo. More accurate 
representations of the stellar and dark matter distributions are possible
 (e.g.\ Mamon \& {\L}okas \cite{mamon05}), but at this stage we wish 
to use identical models to R03 for consistent comparisons.

For the stellar mass distribution, extensive studies of stellar 
dynamics in the inner regions of NGC~3379 
(Kronawitter et al.\ \cite{kronawitter00}; Gebhardt et al.\ \cite{gebhardt00}; 
Cappellari et al.\ \cite{cappellari05}), 
find stellar mass-to-light ratios of $\Upsilon_{\star,\varB}=5$--6 
(in units of $\Upsilon_{\odot,\varB}$).
These models neglect any dark matter found inside $R_{\rm e}$,
so the actual $\Upsilon_{\star,\varB}$ 
may be lower. The orbit models of R03 separate $M(r)$ into the stellar and
dark components, giving $\Upsilon_{\star,\varB}=6$ 
(based primarily on the kinematics of the stars rather than the PNe).
Stellar population synthesis models with a Kroupa IMF
imply $\Upsilon_{\star,\varB}=4\,\!$--7
(Gerhard et al.\ \cite{gerhard01}; Napolitano et al.\ \cite{napolitano05}; 
Cappellari et al.\ \cite{cappellari05}). We adopt $\Upsilon_{\star,\varB}=5.8$ 
as the overall best estimate, and combine this with the observed
light profile to construct the galaxy mass
profile. The resulting stellar mass models for NGC~3379 gives 
$\varv_{\star}=603$~km\,s$^{-1}$ and 
$a_\star=0\farcm3$ ($R_{\rm e}=0\farcm6$) for use in Eq.~(\ref{eq2}) above.

To determine the $\varv_s$ and $r_s$ values for a given $\Lambda$CDM halo, 
we compute the virial mass and radius, $M_{\rm vir}$ and $r_{\rm vir}$, 
and express the model halos in terms of $M_{\rm vir}$ and the 
concentration $c_{\rm vir} \equiv r_{\rm vir}/r_s$. In $\Lambda$CDM, 
there is a well-determined mean relation between these parameters: 
$c_{\rm vir} \simeq 17.9 
(M_{\rm vir}/10^{11} M_{\odot})^{-0.125}$,
with a 68\% log scatter of 0.14
(Bullock et al.\ \cite{bullock01}; Napolitano et al.\ \cite{napolitano05}).
Our approach here differs from R03 in assuming a fluctuation 
amplitude $\sigma_8\,=\,0.9$ rather than 1.0, and in using a virial 
overdensity of 101 rather than 200 -- 
changes that permit less massive inner halos in $\Lambda$CDM.

We then consider whether $\Lambda$CDM halos with various masses 
and concentrations are consistent with the velocity dispersions 
we find for the NGC~3379 GCs. To do this, we calculate the projected 
velocity dispersion of the globular cluster system using a Jeans equation:
\begin{equation}
\sigma^2_{\rm p}(R) = \frac{2}{\Sigma(R)} \int_R^\infty \frac{r {\rm d}r}{\sqrt{r^2-R^2}} \int_r^\infty \frac{j(r') \varv_{\rm c}^2(r') {\rm d}r'}{r'} .
\end{equation}
In making this calculation of the $\sigma_{\rm p}(R)$ expected for
the GC system given a mass model for the galaxy, we assume the following:
1) spherical symmetry;
2) isotropic distribution function and Gaussian LOSVDs;
3) density distribution of tracers $j(r)$ well-known;
4) dynamically negligible gas content;
5) stellar mass-to-light ratio $\Upsilon_{\star}$ well-constrained 
by other techniques.

The spherical assumption is motivated by the roundness of NGC~3379 
(e.g.\ Peletier et al.\ \cite{peletier90}),
the absence of evidence for substantial flattening of the GC 
system of NGC~3379 (RZ04), and the evidence that globular
cluster systems appear to be typically at least as round as their host
galaxies (e.g., Ashman \& Zepf \cite{ashman98}).
Similarly, isotropy is observed for systems where sufficient data are available
(e.g.\ Zepf et al.\ \cite{zepf00}; C\^ot\'e et al.\ \cite{cote03}); our
$\hat{\sigma}_{\rm p}$ fits 
 to the NGC~3379 GC velocities give very similar results to traditional 
$\sigma_{\rm p}$ calculations, indicating that Gaussianity is a good 
approximation. The density distribution we adopt is a Hernquist 
(\cite{hernquist90}) model which matches the $R^{1/4}$ law fitted to the 
NGC~3379 globular clusters (RZ04): $j(r) \propto r^{-1} (r+a)^{-3}$, where
$a=9\farcm8$ ($R_{\rm e}=17\farcm9$).
The mass of hot gas within 100~kpc of NGC~3379 is no more than $\approx 10^{8} 
M_{\odot}$ (David et al.\ \cite{david05}, conservatively assuming a 
linear increase in gas mass from the inner parts), which is a negligible 
fraction of the stellar mass of $M_{\star}\,\approx\,10^{11} M_{\odot}$.

\subsection{Comparison of globular cluster system velocity dispersion 
to models}

\begin{figure}
   \resizebox{\hsize}{!}{\includegraphics[width=9cm]{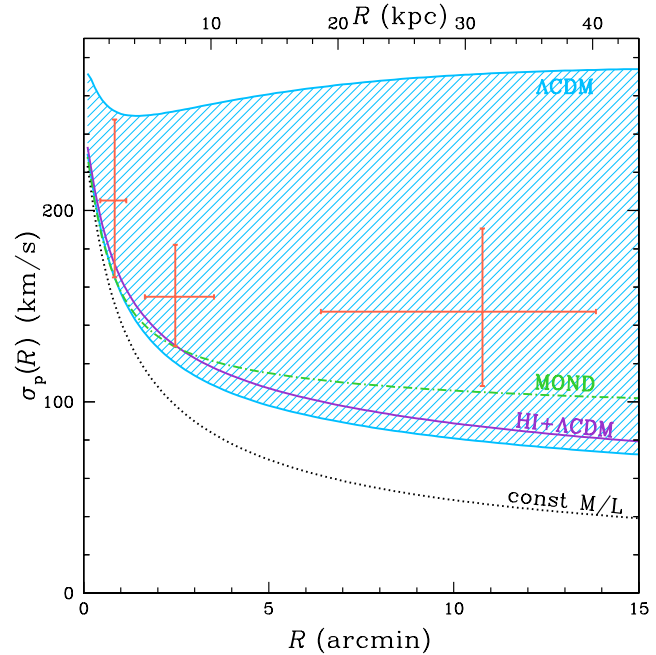}}
      \caption{Model predictions compared to the observed  velocity dispersion 
profile of the NGC~3379 GCs.  Error bars are as described in 
Fig.~\ref{disp}. The hatched region describes profiles for typical 
$\Lambda$CDM halos as described in the text. The solid line shows the
``consensus'' model that matches the \ion{H}{i} data and provides a 
reasonable fit to the GC and PNe results. The  
 dotted line represents a mass traces light model with Newtonian dynamics, 
while the  dot-dashed line is with MOND.}
         \label{disp2}
   \end{figure}

The comparison of the $\sigma_{\rm p}(R)$ of the 
GC data to predictions for various models of the mass distribution 
for NGC~3379 is shown in Fig.~\ref{disp2}.  The GC data are in good agreement
with a wide variety of $\Lambda$CDM halos that have typical concentrations for 
their masses and match the inner stellar data. 
In detail, the range of $\Lambda$CDM halos plotted is based on halos  
which fall within the $1\sigma$ scatter in
the halo mass-concentration relation of Bullock et al.\ (\cite{bullock01}),
with the additional constraints that the stellar mass to dark
halo mass is not lower than typical for galaxy clusters and is
not larger than the cosmic baryon fraction.

A valuable constraint on the halo mass of NGC~3379 is provided by 
observations of a rotating ring of \ion{H}{i} gas at $R \sim 100$~kpc 
(Schneider et al.\ \cite{schneider83}; Schneider \cite{schneider85}).
The positions and velocities of the \ion{H}{i} are consistent with the gas 
being in a Keplerian orbit around the barycentre of NGC~3379 and NGC~3384, 
with about \hbox{$^2\!/_3$} of the mass within 100~kpc belonging to NGC~3379. 
This then gives a mass for NGC~3379 of about 4--5 $\times 10^{11}M_{\odot}$ 
at $R \sim 100$~kpc  and $\Upsilon_\varB \simeq 25$--35.
A $\Lambda$CDM halo model that matches the \ion{H}{i} ring and
is generally consistent with the GC and PN dynamical constraints
is shown in Fig.~\ref{disp2}. This ``consensus'' model predicts a GC 
velocity dispersion that is close to that observed,
with a  preference for  smaller values. 

This ``best-fitting'' $\Lambda$CDM halo is only slightly below
the GC data and has a concentration well within the normal range
for a halo of its mass. The only notable feature of this $\Lambda$CDM halo
is that it gives a fairly small mass-to-light ratio, 
with $\Upsilon_\varB \simeq 35$. Modest mass-to-light ratios for 
early-type galaxies with luminosities around $L^*$ like NGC~3379 may 
be in agreement with expectations from some recent halo occupation models
(e.g.\ van den Bosch et al.\ \cite{vandenbosch03}), although the value 
for NGC~3379 is about a factor of two less than the model average for 
galaxies of its mass.  Another implication
of the inferred mass-to-light ratio is that if the halo has the 
universal baryonic fraction of 0.17 of the concordance model, and 
given the $\Upsilon_\varB$ of about 6 measured in the inner regions, the 
$\Upsilon_\varB$ for the halo would suggest most of the baryons 
have been turned into stars (see Napolitano et al.\ \cite{napolitano05} 
for this calculation for a number of galaxies).

Our data also allow us to assess alternatives to $\Lambda$CDM halos.
Assuming no dark matter and a Newtonian force law, we expect
$\sigma_{\rm p}=98$~km\,s$^{-1}$ at $R=2\farcm5$ and $49$~km\,s$^{-1}$ at 
$10'$. This mass traces light model is clearly ruled out by the data 
(see Fig.~\ref{disp2}).
We also consider MOND, which has been proposed to provide a good
fit to the earlier PNe data (Milgrom \& Sanders \cite{milgrom03}).
The comparison of our observed $\sigma_{\rm p}$ with the
predictions of MOND is shown in Fig.~\ref{disp2}.
The velocity dispersion of the GCs is  
slightly higher than the MOND prediction, although the uncertainties 
  mitigate against an unambiguous rejection of MOND.

A final comparison is to the published velocity dispersions
of NGC~3379 PNe (R03). The reasonable agreement of all
of the constraints with the overall ``best-fit'' mass model
indicates that the GCs and PNe are broadly consistent.
In detail, although the PNe have a $\sigma_{\rm p}$ of about
120~km\,s$^{-1}$ in the radial range of $0\farcm7 \le R \le 4'$
where the PN data overlap with our GC data, compared
to a $\sigma_{\rm p}$ of about 170~km\,s$^{-1}$ for the GCs
in the same region, the two do not give greatly discrepant
masses. This is partly due to radial anisotropy in the PNe
system suggested by orbital modeling of the PN system (R03),
which naturally gives a somewhat lower $\sigma_{\rm p}$ for the same
mass, and partly due to the uncertainties in $\sigma_{\rm p}$ from
the modest number of radial velocities. Furthermore, the GCs
appear to have a shallower spatial profile than the galaxy light
(and by inference the PNe) in the relevant region
(RZ04) which would produce a smaller $\sigma_{\rm p}$ for the
PNe relative to the GCs. If additional data
find a difference in the $\sigma_{\rm p}$ beyond these effects,
several possibilities present themselves. These include
the possibility that the GCs may have modestly tangential
orbits, that orbits may be disturbed by group interactions,
or that NGC~3379 may have a substantial disc component
seen face-on (e.g.\ Capaccioli et al.\ \cite{capaccioli91})
which would be more likely to include PNe than GCs.
The question of the halo mass profile inferred for NGC~3379 has
also been addressed in a recent paper by Dekel et al.\ (\cite{dekel05}).
 They simulated disk galaxy mergers and found that in the 
resulting elliptical galaxies, tracers like PNe have steeply declining 
$\sigma_{\rm p}$ profiles because of radially biased orbits, 
 flattened triaxial structures, and steep spatial profiles. 
They predicted that GCs would be less prone to
these effects and should have flatter dispersion profiles, which is
consistent with our results.

\section{Conclusions \label{sect5}}

We have used the FLAMES multi-fibre spectrograph on the VLT
to obtain radial velocities of globular clusters (GCs) around the Leo I group
galaxies NGC~3379 and NGC~3384. FLAMES allowed us to obtain 
spectra of 42 GCs over a  wide field ($1'\!\la r \la\!14'$) 
with very small uncertainties ($\sim\!10$~km\,s$^{-1}$).

1. Our primary conclusion is that the projected velocity dispersion of
the GC system of NGC~3379 is consistent with that
expected from standard dark matter halos in $\Lambda$CDM cosmologies.
This is based on comparing our observed velocity dispersions for the
GC system over a wide range of radii to the expectation 
from the Jeans equation for the mass profile of $\Lambda$CDM halo 
models and the observed radial profile of the GC system.

2. We find evidence that the GC system of the SB0
galaxy NGC~3384 has a rotation similar to that of the stellar
disc and planetary nebula system of this galaxy. This result
suggests that significant rotation in the globular cluster systems 
of disc galaxies may not be rare.

3. We find a colour gradient in our spectroscopic sample
such that the fraction of red, metal-rich GCs decreases
from small to large radii, in agreement with photometric
studies. Even with this colour gradient, we also find some
red GCs at large distances from the centre of their
host galaxy, which may place interesting constraints on models
for the origin of metal-rich globular clusters.

\begin{acknowledgements}
GB is supported by the
Secretar\'\i a de Estado de Universidades e Investigaci\'on with additional
support by DGI (Spain) AYA 2002-03338 and Junta de Andaluc\'ia TIC-114.
SEZ acknowledges support for this work in part from NSF AST-0406891 
and from the Michigan State University Foundation. 
AJR is supported by the FONDAP Center for Astrophysics Conicyt 15010003.
KLR is supported by an NSF Astronomy and Astrophysics Postdoctoral Fellowship 
under award AST-0302095. 
\end{acknowledgements}

\appendix
\section{Fitting dispersion profiles to discrete velocity data}
Radial binning or smoothing is not ideal for characterising data and
testing models, as the grouping of data discards useful information
about the radial locations $R_i$ and may skew the result
by assuming $\sigma_{\rm p}$ is constant within each bin.
As an alternative, we consider a maximum-likelihood technique for fitting 
discrete velocity data to local velocity dispersions (see Ciardullo et al. 
\cite{ciardullo93}; Saglia et al. \cite{saglia00}).
This technique allows one to fit data to a model
for $\sigma_{\rm p}(R)$, which may be e.g.\ a constant dispersion, a
  power-law, or the prediction of a Jeans model.

We convolve a Gaussian line-of-sight velocity distribution (LOSVD)
with a Gaussian measurement function $\varv_i\pm\delta \varv_i$ to derive
the likelihood of a given measurement $\varv_i$ given a model dispersion
$\sigma_{\rm p}(R_i)$:
\begin{equation}
{\cal L}_i = \frac{1}{\sqrt{\sigma_{\rm p}^2+(\delta \varv_i)^2}}\exp\left\{\frac{(\varv_i-\varv_0)^2}{2\left[\sigma_{\rm p}^2+(\delta \varv_i)^2\right]}\right\} ,
 \end{equation}
which can be converted into a $\chi^2$ statistic:
\begin{equation}
\chi^2 = \sum_{i=1}^{i=N}\left\{\frac{(\varv_i-\varv_0)^2}{\sigma_{\rm p}^2+(\delta \varv_i)^2} + \ln \left[\sigma_{\rm p}^2+(\delta \varv_i)^2\right] \right\} .
\end{equation}
This method assumes Gaussian LOSVDs, so strictly speaking, we are
computing the best-fit Gaussian parameter $\hat{\sigma}_{\rm p}$
rather than the true second-moment $\sigma_{\rm p}$. For near-isotropic
systems, $\hat{\sigma}_{\rm p}\simeq \sigma_{\rm p}$.

To apply this to our globular cluster data for NGC~3379, we characterise 
the dispersion data by a scale-free power-law profile:
\begin{equation}
\sigma_{\rm p}^2(R) = \sigma_0^2 (R/R_0)^{-\gamma_{\rm p}} ,
\end{equation}
where $R_0=2\farcm2$. We find 
$\sigma_0=168^{+21}_{-19}$~km\,s$^{-1}$ and $\gamma_{\rm p}=0.25\pm0.25$ 
(see Fig.~\ref{disp}).
This result is scarcely affected by including the four uncertain clusters;
on the other hand, adopting a mean velocity of $904$~km\,s$^{-1}$ would 
mean $\sigma_0=177^{+21}_{-21}$~km\,s$^{-1}$ and 
$\gamma_{\rm p}=0.20\pm0.25$.

This dispersion estimator is checked using Monte Carlo simulations.  A
small correction for estimator bias is derived using the difference
between the simulations' input and median output.  The (doubly-debiassed)
68\% scatter of the simulations gives the likely range of the dispersion.

\Online

\begin{table*}
\caption{Heliocentric radial velocities $\varv$ (in km\,s$^{-1}$)
of FLAMES candidate globular clusters in Leo I. The first 34 (``Class A'',
 ordered by $\varV\!$-band magnitude), are certain; the next 8 (``Class B'') 
are probable; the final 8 (``Class C'') are possible. Object ID and photometry 
are from Rhode \& Zepf (\cite{rhode04}) --~the ``{\sf ad}'' prefixed 
objects have been added by hand to their initial list of GC candidates.  
$h_{_{\rm CCF}}$, $w_{_{\rm CCF}}$ and $\mathcal{R}_{_{\rm CCF}}$ are 
the height, width (in pixels) and Tonry \& Davis (\cite{tonry79}) 
quality coefficient of the Gaussian fit to
the cross-correlation function peak. The inferred {\texttt{fxcor}} 
velocities and their estimated errors 
are the median values over the \'ELODIE templates.
}
\label{globulars}
\centering
\begin{tabular}{lccccc rrr rl l}
\hline\hline
Id$_{\rm{RZ04}}$ & $\alpha\,$({\it J2000.0}) & $\delta\ \,$({\it J2000.0}) & $\varV$ & $\varB-\varV$ & $\varV-\varR$ & $h_{_{\rm CCF}}$&$w_{_{\rm CCF}}$&$\mathcal{R}_{_{\rm CCF}}$&$\varv$ &$\delta\varv$&Notes\\
\hline
{\sf ad1102 } & 10:48:10.29 & +12:40:57.9 & 19.28 & 0.75 & 0.50 & 0.46 & 71  & 23.8 & 920 &  3  & Class A\\
{\sf ad1481 } & 10:47:52.08 & +12:21:45.2 & 19.39 & 0.52 & 0.43 & 0.16 & 61  & 7.0  & 1105&  6  & Class A\\
{\sf gc764  } & 10:47:51.07 & +12:35:49.1 & 20.00 & 0.72 & 0.52 & 0.37 & 96  & 18.7 & 776 &  4  & Class A\\
{\sf gc960  } & 10:47:39.25 & +12:32:25.6 & 20.06 & 0.81 & 0.49 & 0.29 & 87  & 14.9 & 899 &  5  & Class A\\
{\sf gc948  } & 10:47:40.34 & +12:35:54.9 & 20.25 & 0.66 & 0.38 & 0.16 & 89  & 6.8  & 1264&  9  & Class A\\
{\sf gc670  } & 10:47:58.08 & +12:28:11.2 & 20.37 & 0.72 & 0.44 & 0.15 & 65  & 7.4  & 921 &  6  & Class A\\
{\sf gc682  } & 10:47:56.72 & +12:33:26.4 & 20.47 & 0.63 & 0.48 & 0.17 & 67  & 8.1  & 1137&  6  & Class A\\
{\sf gc984  } & 10:47:37.80 & +12:31:52.7 & 20.86 & 0.82 & 0.49 & 0.18 & 63  & 7.8  & 1130&  5  & Class A\\
{\sf gc573  } & 10:48:07.19 & +12:37:55.1 & 20.94 & 0.95 & 0.57 & 0.28 & 68  & 13.0 & 752 &  4  & Class A\\
{\sf gc820  } & 10:47:48.18 & +12:35:44.9 & 20.94 & 0.75 & 0.49 & 0.24 & 100 & 9.9  & 889 &  6  & Class A\\
{\sf gc709  } & 10:47:54.19 & +12:36:31.5 & 20.97 & 0.72 & 0.50 & 0.27 & 73  & 13.0 & 954 &  4  & Class A\\
{\sf gc634  } & 10:48:02.13 & +12:35:57.3 & 20.97 & 0.77 & 0.46 & 0.16 & 68  & 7.0  & 950 &  6  & Class A\\
{\sf gc449  } & 10:48:16.20 & +12:38:30.6 & 20.97 & 0.80 & 0.54 & 0.36 & 80  & 15.8 & 672 &  4  & Class A\\
{\sf gc645  } & 10:48:00.52 & +12:32:46.9 & 21.02 & 0.78 & 0.48 & 0.18 & 73  & 7.5  & 654 &  6  & Class A\\
{\sf gc185  } & 10:48:38.75 & +12:27:09.3 & 21.09 & 0.64 & 0.37 & 0.12 & 77  & 4.8  & 964 &  10 & Class A\\
{\sf gc830  } & 10:47:47.71 & +12:34:14.5 & 21.11 & 0.85 & 0.59 & 0.31 & 123 & 16.4 & 926 &  5  & Class A\\
{\sf gc683  } & 10:47:56.61 & +12:26:30.2 & 21.14 & 0.70 & 0.39 & 0.10 & 101 & 4.4  & 1087&  14 & Class A\\
{\sf gc442  } & 10:48:16.81 & +12:37:04.0 & 21.15 & 0.74 & 0.51 & 0.19 & 134 & 7.7  & 878 &  11 & Class A\\
{\sf gc368  } & 10:48:22.19 & +12:44:00.3 & 21.15 & 0.82 & 0.44 & 0.09 & 75  & 4.6  & 1098&  10 & Class A\\
{\sf gc925  } & 10:47:41.87 & +12:33:31.0 & 21.18 & 0.99 & 0.56 & 0.29 & 71  & 13.2 & 1048&  4  & Class A\\
{\sf gc571  } & 10:48:07.40 & +12:39:24.1 & 21.25 & 0.73 & 0.50 & 0.13 & 79  & 5.9  & 792 &  9  & Class A\\
{\sf gc771  } & 10:47:50.62 & +12:35:31.9 & 21.26 & 0.68 & 0.52 & 0.14 & 90  & 6.0  & 758 &  9  & Class A\\
{\sf gc741  } & 10:47:52.60 & +12:35:59.6 & 21.27 & 0.97 & 0.59 & 0.24 & 76  & 9.6  & 1188&  6  & Class A\\
{\sf gc358  } & 10:48:22.79 & +12:38:44.2 & 21.35 & 0.59 & 0.46 & 0.12 & 112 & 5.6  & 577 &  13 & Class A\\
{\sf gc774  } & 10:47:50.40 & +12:33:47.7 & 21.36 & 0.68 & 0.50 & 0.13 & 94  & 5.3  & 656 &  9  & Class A\\
{\sf gc482  } & 10:48:13.88 & +12:37:14.1 & 21.59 & 1.00 & 0.63 & 0.31 & 137 & 17.6 & 908 &  6  & Class A\\
{\sf gc351  } & 10:48:23.34 & +12:36:57.7 & 21.60 & 1.02 & 0.49 & 0.09 & 109 & 3.9  & 739 &  17 & Class A\\
{\sf ad1021 } & 10:48:13.33 & +12:42:24.8 & 21.71 & 0.99 & 0.52 & 0.12 & 56  & 5.3  & 752 &  7  & Class A\\
{\sf gc728  } & 10:47:53.42 & +12:34:12.2 & 21.83 & 0.81 & 0.44 & 0.13 & 134 & 5.2  & 978 &  15 & Class A\\
{\sf gc740  } & 10:47:52.65 & +12:33:37.9 & 21.85 & 1.07 & 0.52 & 0.14 & 73  & 6.2  & 889 &  7  & Class A\\
{\sf gc461  } & 10:48:15.36 & +12:36:58.0 & 21.87 & 0.71 & 0.45 & 0.18 & 216 & 9.4  & 847 &  17 & Class A\\
{\sf gc454  } & 10:48:15.73 & +12:35:54.2 & 21.89 & 1.06 & 0.62 & 0.13 & 79  & 6.1  & 874 &  8  & Class A\\
{\sf gc387  } & 10:48:20.06 & +12:37:15.5 & 21.98 & 0.83 & 0.44 & 0.12 & 87  & 5.4  & 704 &  11 & Class A\\
{\sf gc839  } & 10:47:47.19 & +12:34:04.6 & 21.98 & 0.69 & 0.46 & 0.17 & 152 & 6.7  & 1064&  13 & Class A\\
\hline
{\sf gc756  }& 10:47:51.58  & +12:45:40.2 & 20.87 & 0.58 & 0.47 & 0.09 & 163 & 3.7  & 725 &  19 & Class B\\
{\sf gc719  }& 10:47:53.83  & +12:33:46.2 & 21.51 & 0.65 & 0.52 & 0.10 & 79  & 4.4  & 740 &  10 & Class B\\
{\sf gc958  }& 10:47:39.34  & +12:33:59.1 & 21.57 & 0.74 & 0.42 & 0.09 & 58  & 4.4  & 951 &  8  & Class B\\
{\sf gc946  }& 10:47:40.46  & +12:31:36.4 & 21.67 & 0.65 & 0.36 & 0.09 & 44  & 4.3  & 739 &  6  & Class B\\
{\sf gc971  }& 10:47:38.75  & +12:36:47.8 & 21.78 & 0.83 & 0.47 & 0.07 & 162 & 4.4  & 1027&  13 & Class B\\
{\sf gc949  }& 10:47:40.33  & +12:37:24.1 & 21.81 & 0.69 & 0.43 & 0.06 & 107 & 3.0  & 933 &  27 & Class B\\
{\sf gc581  }& 10:48:06.60  & +12:30:58.5 & 21.82 & 0.98 & 0.58 & 0.08 & 71  & 3.9  & 1064&  11 & Class B\\
{\sf gc923  }& 10:47:41.91  & +12:36:11.2 & 21.97 & 0.75 & 0.57 & 0.08 & 129 & 4.0  & 999 &  14 & Class B\\
\hline
{\sf gc483  }& 10:48:13.81  & +12:31:05.0 & 20.34 & 0.60 & 0.41 & 0.08 & 63  & 3.5  & 667 &  13 & Class C\\
{\sf gc857  }& 10:47:46.02  & +12:31:17.9 & 21.14 & 0.77 & 0.42 & 0.08 & 82  & 3.7  & 897 &  12 & Class C\\
{\sf gc607  }& 10:48:04.64  & +12:25:15.3 & 21.22 & 0.64 & 0.34 & 0.07 & 172 & 3.1  & 1237&  18 & Class C\\
{\sf gc626  }& 10:48:02.87  & +12:45:48.1 & 21.62 & 0.79 & 0.56 & 0.10 & 80  & 4.3  & 771 &  9  & Class C\\
{\sf gc901  }& 10:47:43.42  & +12:33:33.7 & 21.76 & 0.54 & 0.35 & 0.06 & 139 & 3.7  & 632 &  11 & Class C\\
{\sf gc143  }& 10:48:42.52  & +12:35:15.0 & 21.77 & 0.67 & 0.35 & 0.10 & 50  & 4.8  & 1407&  7  & Class C\\
{\sf gc661  }& 10:47:58.86  & +12:33:54.5 & 21.89 & 0.84 & 0.39 & 0.09 & 97  & 4.6  & 954 &  8  & Class C\\
{\sf gc538  }& 10:48:10.74  & +12:32:42.6 & 21.89 & 1.08 & 0.54 & 0.11 & 60  & 4.2  & 877 &  10 & Class C\\ 
\hline
\end{tabular}
\end{table*}

\begin{table*}
\caption{Heliocentric radial velocities $\varv$ (in km\,s$^{-1}$)
of FLAMES candidate stars.  The first 32 (``Class A''), 
ordered by $\varV\!$-band magnitude, are certain; the next 4 (``Class B'') 
are probable; the final one (``Class C'') is a possible star.
Object ID and photometry from Rhode \& Zepf (\cite{rhode04}). $h_{_{\rm CCF}}$,
  $w_{_{\rm CCF}}$ and $\mathcal{R}_{_{\rm CCF}}$ are the height,
width (in pixels) and Tonry \& Davis (\cite{tonry79}) quality 
coefficient of the Gaussian fit to the cross-correlation function peak. 
The inferred {\texttt{fxcor}} radial velocities
 and their estimated errors are the median values over the 
\'ELODIE templates.}
\label{stars}
\centering
\begin{tabular}{lccccc c rrr ll l}
\hline\hline
Id$_{\rm{RZ04}}$ & $\alpha\,$({\it J2000.0}) & $\delta\ $({\it J2000.0}) & $\varV$ & $\varB-\varV$ & $\varV-\varR$ & $h_{_{\rm CCF}}$&$w_{_{\rm CCF}} $&$\mathcal{R}_{_{\rm CCF}}$&$\varv$ &$\delta\varv$&Notes\\
\hline 
{\sf gc139 } & 10:48:43.10 & +12:38:05.6 & 19.15 & 0.68 & 0.42 &  0.54 & 83 &  27.6&  27  &  2  & Class A  \\ 
{\sf gc737 } & 10:47:52.80 & +12:44:49.4 & 19.16 & 0.70 & 0.43 &  0.61 & 77 &  33.1&  51  &  2  & Class A  \\ 
{\sf gc516 } & 10:48:12.05 & +12:34:27.0 & 19.58 & 0.67 & 0.40 &  0.38 & 88 &  19.4&  78  &  3  & Class A  \\ 
{\sf ad945 } & 10:48:15.83 & +12:24:19.3 & 19.64 & 0.64 & 0.38 &  0.28 & 69 &  14.3&  56  &  4  & Class A  \\ 
{\sf gc1272} & 10:47:14.12 & +12:33:00.0 & 19.71 & 0.71 & 0.41 &  0.32 & 76 &  14.7&  129 &  4  & Class A  \\ 
{\sf gc660 } & 10:47:58.88 & +12:29:49.5 & 19.83 & 0.98 & 0.59 &  0.50 & 80 &  22.0&  129 &  3  & Class A  \\ 
{\sf ad497 } & 10:48:35.68 & +12:24:43.2 & 19.93 & 0.69 & 0.38 &  0.19 & 70 &  8.2 &  209 &  6  & Class A  \\ 
{\sf ad411 } & 10:48:39.25 & +12:25:51.2 & 19.99 & 0.56 & 0.34 &  0.15 & 70 &  6.7 &  $-$93 &  7  & Class A  \\ 
{\sf gc290 } & 10:48:28.21 & +12:41:15.5 & 20.00 & 0.76 & 0.43 &  0.18 & 91 &  8.8 &  158 &  7  & Class A  \\ 
{\sf gc220 } & 10:48:35.45 & +12:29:49.1 & 20.03 & 0.68 & 0.38 &  0.26 & 75 &  12.3&  89  &  5  & Class A  \\ 
{\sf gc499 } & 10:48:13.15 & +12:33:49.9 & 20.04 & 0.79 & 0.44 &  0.21 & 64 &  10.5&  213 &  4  & Class A  \\ 
{\sf gc950 } & 10:47:40.33 & +12:35:23.1 & 20.10 & 0.59 & 0.39 &  0.20 & 93 &  9.5 &  42  &  7  & Class A  \\ 
{\sf gc277 } & 10:48:29.49 & +12:35:20.0 & 20.16 & 0.66 & 0.37 &  0.16 & 102&  4.4 &  $-$34 &  14 & Class A  \\ 
{\sf gc605 } & 10:48:04.73 & +12:38:35.0 & 20.18 & 0.61 & 0.40 &  0.10 & 78 &  5.0 &  36  &  11 & Class A  \\ 
{\sf gc236 } & 10:48:33.54 & +12:43:04.8 & 20.20 & 0.79 & 0.48 &  0.30 & 75 &  13.6&  138 &  4  & Class A  \\ 
{\sf gc814 } & 10:47:48.42 & +12:44:49.0 & 20.24 & 0.64 & 0.41 &  0.14 & 65 &  6.4 &  $-$15 &  7  & Class A  \\ 
{\sf gc988 } & 10:47:37.30 & +12:33:33.9 & 20.25 & 0.92 & 0.56 &  0.28 & 106&  12.3&  102 &  6  & Class A  \\ 
{\sf gc980 } & 10:47:38.05 & +12:39:31.1 & 20.39 & 0.69 & 0.39 &  0.16 & 94 &  7.7 &  93  &  9  & Class A  \\ 
{\sf gc160 } & 10:48:41.07 & +12:29:32.0 & 20.44 & 0.60 & 0.32 &  0.12 & 57 &  5.8 &  86  &  7  & Class A  \\ 
{\sf gc1066} & 10:47:31.54 & +12:40:20.4 & 20.73 & 0.94 & 0.45 &  0.22 & 82 &  10.8&  $-$61 &  5  & Class A  \\ 
{\sf gc542 } & 10:48:10.60 & +12:45:04.5 & 20.75 & 0.63 & 0.31 &  0.10 & 59 &  4.7 &  31  &  8  & Class A  \\ 
{\sf gc443 } & 10:48:16.80 & +12:26:35.6 & 20.94 & 0.76 & 0.37 &  0.16 & 91 &  5.0 &  $-$57 &  14 & Class A  \\ 
{\sf gc480 } & 10:48:14.32 & +12:23:22.8 & 21.06 & 0.87 & 0.49 &  0.18 & 59 &  8.2 &  153 &  5  & Class A  \\ 
{\sf gc451 } & 10:48:15.85 & +12:39:46.2 & 21.06 & 0.74 & 0.43 &  0.13 & 63 &  5.4 &  204 &  8  & Class A  \\ 
{\sf gc940 } & 10:47:40.73 & +12:39:45.4 & 21.14 & 1.01 & 0.48 &  0.14 & 53 &  7.0 &  332 &  6  & Class A  \\ 
{\sf gc635 } & 10:48:02.07 & +12:32:44.3 & 21.17 & 0.80 & 0.45 &  0.11 & 88 &  5.6 &  $-$78 &  10 & Class A  \\ 
{\sf gc493 } & 10:48:13.55 & +12:39:20.7 & 21.19 & 0.98 & 0.57 &  0.26 & 73 &  10.5&  133 &  6  & Class A  \\ 
{\sf gc541 } & 10:48:10.71 & +12:35:05.0 & 21.21 & 0.73 & 0.34 &  0.09 & 80 &  4.2 &  $-$29 &  12 & Class A  \\ 
{\sf gc311 } & 10:48:26.34 & +12:38:32.7 & 21.34 & 0.83 & 0.51 &  0.12 & 98 &  6.2 &  $-$102&  10 & Class A  \\ 
{\sf gc981 } & 10:47:37.98 & +12:40:53.6 & 21.34 & 0.80 & 0.45 &  0.16 & 80 &  7.4 &  280 &  8  & Class A  \\ 
{\sf gc1089} & 10:47:29.21 & +12:42:16.2 & 21.41 & 0.76 & 0.50 &  0.09 & 59 &  4.2 &  $-$58 &  8  & Class A  \\ 
{\sf gc904 } & 10:47:43.21 & +12:36:47.4 & 21.53 & 0.84 & 0.46 &  0.16 & 83 &  8.0 &  147 &  7  & Class A  \\ 
\hline
{\sf ad508 } & 10:48:35.04 & +12:34:10.6 & 20.69 & 0.91 & 0.46 &  0.09 & 79 &  4.1 &  12  & 12  & Class B  \\
{\sf gc428 } & 10:48:17.76 & +12:40:34.7 & 21.09 & 0.57 & 0.30 &  0.07 & 62 &  3.4 &  112 & 11  & Class B  \\
{\sf gc1198} & 10:47:20.28 & +12:36:24.5 & 21.41 & 0.79 & 0.45 &  0.08 & 49 &  3.9 &  16  & 10  & Class B  \\
{\sf gc599 } & 10:48:04.86 & +12:34:08.7 & 21.92 & 0.55 & 0.39 &  0.10 & 100&  4.5 &  12  & 14  & Class B  \\
\hline
{\sf gc398 } & 10:48:19.55 & +12:28:13.2 & 21.22 & 0.70 & 0.34 &  0.07 & 78 &  3.4 &  229 & 14  & Class C  \\
\hline
\end{tabular}
\end{table*}

\begin{table*}
\caption{The 21 unclassified FLAMES targets, ordered by $\varV\!$-band 
magnitude. Object ID and photometry are from Rhode \& Zepf (\cite{rhode04}).}
\label{galaxies}
\centering
\begin{tabular}{l c c c c c l}
\hline\hline
Id$_{\rm{RZ04}}$&$\alpha\,({\it J2000.0})$&$\delta~\,({\it J2000.0})$&$\varV$&$\varB-\varV$&$\varV-\varR$&Notes\\
\hline
{\sf gc465 } &10:48:15.18 &+12:41:03.3 &20.55 &0.73 &0.46& CCD bad columns!\\
{\sf gc749 } &10:47:52.12 &+12:26:52.1 &20.91 &0.72 &0.40&\\
{\sf gc871 } &10:47:45.28 &+12:39:59.8 &21.10 &0.62 &0.30&\\
{\sf gc1176} &10:47:22.40 &+12:30:44.1 &21.15 &0.69 &0.32&\\
{\sf gc1080} &10:47:30.38 &+12:40:53.4 &21.16 &0.73 &0.37&\\
{\sf gc970 } &10:47:38.84 &+12:44:27.6 &21.19 &0.53 &0.39&\\
{\sf gc614 } &10:48:03.92 &+12:39:46.4 &21.43 &1.06 &0.48&\\
{\sf gc873 } &10:47:45.16 &+12:33:17.7 &21.62 &0.66 &0.46&\\
{\sf ad1106} &10:48:10.08 &+12:24:08.9 &21.64 &0.73 &0.40& [\ion{O}{ii}] at $z \sim$ 0.40\\
{\sf gc718 } &10:47:53.88 &+12:43:16.6 &21.66 &0.52 &0.42&\\
{\sf gc511 } &10:48:12.27 &+12:27:42.2 &21.70 &0.76 &0.38&\\
{\sf gc803 } &10:47:48.84 &+12:45:03.2 &21.74 &1.04 &0.64&\\
{\sf gc1075} &10:47:30.95 &+12:31:06.6 &21.74 &0.84 &0.49&\\
{\sf gc738 } &10:47:52.79 &+12:27:27.7 &21.76 &0.91 &0.50& Emission lines?
\\
{\sf gc849 } &10:47:46.57 &+12:42:35.9 &21.78 &0.82 &0.47& [\ion{O}{ii}] at $z \sim$ 0.38\\
{\sf gc983 } &10:47:37.91 &+12:35:11.8 &21.78 &0.82 &0.50&\\
{\sf ad697 } &10:48:25.26 &+12:39:11.8 &21.82 &0.94 &0.50&\\
{\sf gc292 } &10:48:28.21 &+12:38:28.6 &21.84 &0.64 &0.42&\\
{\sf gc419 } &10:48:18.12 &+12:34:41.9 &21.84 &0.90 &0.45&\\
{\sf gc344 } &10:48:23.74 &+12:29:17.9 &21.93 &0.69 &0.45& [\ion{O}{ii}] at $z \sim$ 0.40\\
{\sf gc600 } &10:48:04.85 &+12:35:38.3 &21.93 &0.77 &0.48&\\
\hline
\end{tabular}
\end{table*}
\end{document}